# Pearling in cells: A clue to understanding cell shape

Roy Bar-Ziv, Tsvi Tlusty, Elisha Moses, Samuel A. Safran, and Alexander Bershadsky

Departments of *Physics of Complex Systems, ‡Materials and Interfaces, and ¶Molecular Cell Biology, Weizmann Institute of Science, Rehovot 76100, Israel

**ABSTRACT** Gradual disruption of the actin cytoskeleton induces a series of structural shape changes in cells leading to a transformation of cylindrical cell extensions into a periodic chain of "pearls." Quantitative measurements of the pearling instability give a square-root behavior for the wavelength as a function of drug concentration. We present a theory that explains these observations in terms of the interplay between rigidity of the submembranous actin shell and tension that is induced by boundary conditions set by adhesion points. The theory allows estimation of the rigidity and thickness of this supporting shell. The same theoretical considerations explain the shape of nonadherent edges in the general case of untreated cells.

The parameters that determine cell shape are generally well known: adhesion to the substrate, membrane elasticity, and cytoskeleton mechanics (1–4). Yet, a quantitative and predictive model for cell shape is still lacking. In particular, the interplay of the cytoskeleton rigidity with the tension produced by constraints subjected on the cell by adhesion points has not been accounted for in a quantitative framework.

The importance of actin in the majority of processes of cell morphogenesis leads us to probe in detail the changes arising on gradual disruption of the actin cytoskeleton, by using the drug latrunculin A (LatA). LatA is known to bind monomeric actin in a 1:1 complex, sequestering it and thereby allowing control of the level of polymerized actin by varying the drug concentration (5–7). We show here that a gradual increase in LatA concentration leads first to arborization (the formation of numerous radial tubular protrusions), a phenomenon that has previously been described for other drugs that disrupt the actin cytoskeleton. Further increase of the LatA concentration induces an instability of these tubes, converting them into a chain of pearls. This "pearling" is a general phenomenon of tubes under tension that can be found in a wide variety of physical systems (8), including phospholipid bilayers (9).

We quantify the dynamics of the instability in detail and present a theory that explains these phenomena in terms of the competition between the tension in the membrane and the rigidity of the actin cytoskeleton that opposes it. It is interesting that the same theory can quantitatively explain the shape of adherent cells during the arborization process. Moreover, the theory describes the shape of both cells with disrupted actin cytoskeleton, as well as untreated, normally adhering cells. Thus, a simple description in terms of rigidity, sustained by a $h \approx 1 \mu m$ thick actin shell underlying a tense lipid bilayer (10, 11) can explain major features of cell morphogenesis. Comparison of theory and experiment allows estimation of the actin-shell thickness and elastic moduli in both arborized and pearled states and provides a quantitative measure of the rigidity as it is reduced by LatA.

## Observations

SVT2 cells (12) were plated on coverslips and treated with increasing concentration of LatA (0.08–40 $\mu$M). Untreated cells are polygonal, with lamellipodia and protrusions concentrated in one or a few locations on the cell periphery (Fig. 1a). Their edges are curved, and the protrusions are at the vertices. This morphology is disrupted by the addition of LatA in two stages. First, the outer envelope shrinks to form a round cell body enclosing the nucleus and most of the cytoplasm, leaving radial tubular protrusions still attached to adhesion points (a "sun" with "rays"). In the second stage, some of the tubes destabilize, transforming into the pearled state characteristic of tense cylinders. Complete recovery of cell shape occurs on removal of LatA (data not shown).

Careful examination of the cell morphology during arborization shows that the shape is produced by stretching the cell surface between adhesion points creating curved edges (13–15). As the concentration increases, the hanging surface droops, and its curvature increases (Fig. 1 b and c). The part that is near the cell body is well approximated by a circular shape. With progressively higher concentrations of LatA, the cell is rounded, and the curvature of the surface between protrusions changes sign (Fig. 1d). At this stage, instability of the cylinders often occurs (Fig. 1e).

Electron microscopy of the tubular extensions taken after treatment with the drug confirmed that these cylindrical regions contain microtubule bundles. Real-time observations of dynamic behavior, such as buckling within the tubular protrusions, may be attributed to these microtubules because they grow and push against the membrane enveloping the cell. However, the pearling instability cannot be inhibited by total disruption of the microtubules, which is achieved by pretreatment of the cells with $10 \mu M$ nocodazole (data not shown).

The Rayleigh-like pearling instability of these extensions is depicted in Fig. 2. Pearling appears as a finite-amplitude, peristaltic modulation of the cylinder, characterized by a well-defined wavelength. The shape in the highly nonlinear state of the instability is that of isolated spherical pearls of membrane connected by thin tubular sections. The instability occurs predominantly in thinner tubes.

Observing the dynamics of the instability, we found that as the modulation progresses in time, an initially long wavelength can evolve into a final shorter one. The tension initiating the instability often is induced by elongation, related either to motion of the cell as the adhesion sites remain fixed or to the motion of microtubule bundles inside the cylindrical extension. Some tubes are unstable to the pearling instability even with no drug added, but this is a relatively rare event that occurs in existing long tubular protrusions, for example, those that connect neighboring cells.

To quantify the characteristics of the instability we treated the cells with LatA at various concentrations for 1 h and then fixed them. Cells were fixed by 3% paraformaldehyde in PBS;

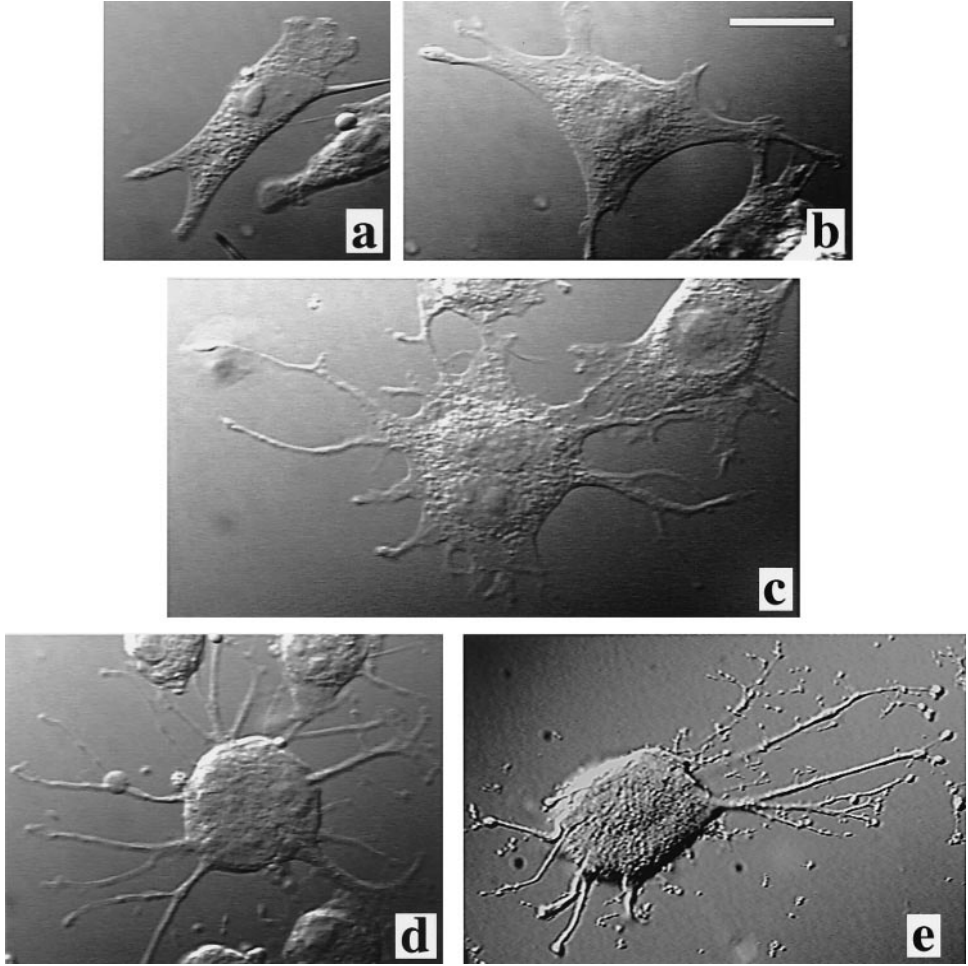

FIG. 1. Microscope images taken in differential interference contrast (×63, oil-immersion objective) of SVT2 cells in varying degree of arborization after treatment with LatA. Concentrations $\Phi$ of the drug are (*a*) control $\Phi = 0$; (*b*) $\Phi = 0.62\ \mu$M; (*c*) $\Phi = 1.25\mu$M; (*d*) $\Phi = 2.5\ \mu$M; and (*e*) $\Phi = 10\ \mu$M. *a* and *b* are most probably mononuclear, whereas in c-e we chose more spherical, multinuclear cells. This illustrates the arborization by enhancing both the symmetry and the number of arbors. (Bar = 20 $\mu$m.)

this fixation did not change the morphology of the pearled tubular protrusions. Measuring the percentage of fixed cells that have undergone pearling shows that this occurs in a range of LatA concentration generally higher than that required for arborization. As the drug concentration, $\Phi$, is increased, the probability for pearling also rises. Fig. 3 (*Lower*) shows a fit of the measured probability to a normal distribution (i.e., this curve is its integral—an error function) with a mean of $\Phi_c = 2.5 \pm 1\ \mu$M and a variance that is of the same order as $\Phi_c$.

The wavelength $\lambda$ also shows a clear trend with $\Phi$. Fig. 3 (*Upper*) shows the measurement of the dimensionless wavenumber, $k \equiv qR_0 = 2\pi R_0/\lambda$, that characterizes the instability at different drug concentrations, averaged over many cells. Varying $\Phi$ by over two decades, we obtain a clear power law behavior $k = (0.42 \pm 0.03) \times \Phi^{0.51 \pm 0.04}$ over most of the range covered.

We also observed the characteristic change of shape associated with actin disruption, as well as the pearling of tubular extensions on treatment with cytochalasin D (results not shown). However, the quantitative relation of the wavelength to the concentration of cytochalasin was not apparent in this case. The complicated mechanism by which cytochalasin acts to disrupt the actin cytoskeleton creates a nonlinear relationship between drug concentration and the extent of actin polymerization (16–18). The linear relation of drug concentration to F actin disruption seems to be a unique property of LatA, which is also the only drug whose specificity for actin has been shown genetically (6, 7).

## Theory

Whereas the same forces of tension and elastic stress act on the cell independent of its shape, the changes in geometry modify its stability properties. The driving force for the pearling instability is the tension in the cell. Typical measured values of this effective tension in untreated cells are $\sigma \simeq 4 \times 10^{-2}$ erg/cm$^2$ (19–21).

The precise origin of tension is an interesting question but is not crucial for our model. Most probably it lies in passive elements such as the membrane, coming from the boundary constraints set by the adhesion points that link the cell to the substrate. Alternatively, active actomyosin contractility could contribute, but in this case the effective tension energy $U\sigma$ would only be redefined. In multicellular tubes like blood vessels, the activation of actomyosin contractility by the drug angiotension II was shown to induce the pearling instability (22). LatA is highly specific and should not affect the actin-myosin interaction directly. Its addition should therefore not affect this active component until the actin filaments get too short as a result of LatA-induced depolymerization.

We addressed the question of passive versus active tension production in a preliminary experiment by using 2,3-butanedione monoxime (BDM, 10–30 mM), a known inhibitor of interactions of actin with myosin (23). We observed LatA-induced pearling even at the highest concentration of BDM (data not shown), suggesting the primary importance of the passive component.

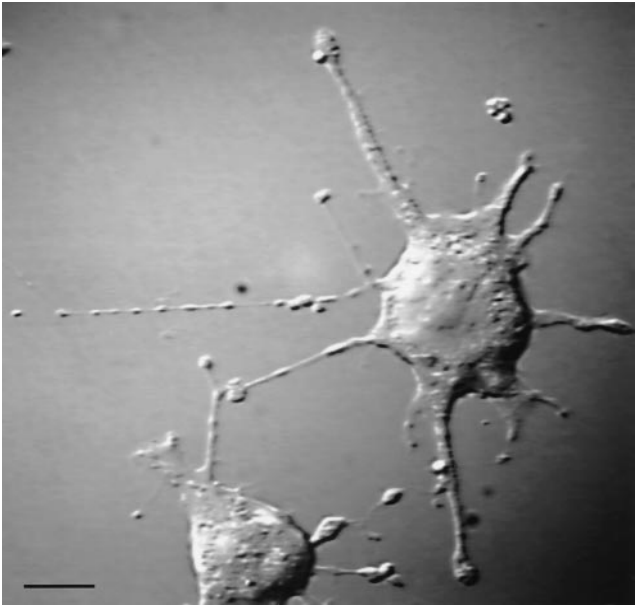

FIG. 2. Pearling in an SVT2 cell treated with 2.5 μM LatA and fixed in paraformaldehyde. All samples included in the statistical analysis were fixed and then viewed in a differential interference contrast microscope by using a computer-enhanced video system. Dynamical observations were conducted in a temperature-stabilized dish held at 37°C. (Bar = 10 μm.)

The force that resists pearling is caused by the shear rigidity in the solid-like actin cortex, which resides beneath the membrane (10, 11). The shear strength of an actin shell reflects its response to a deforming strain and is characterized by its Young's modulus $E$. In an actin gel, $E$ increases with the density of filaments and the number links between them [typically, $E \approx 10^2$–$10^3$ Pa for highly cross-linked actin gels (24–27)]. Increasing the LatA concentration $\Phi$ decreases the amount of polymerized actin in a linear fashion (5) and consequently reduces its Young's modulus $E$. Our measurements are performed far enough from the percolation point (where the network is so disrupted that its macroscopic shear rigidity falls to zero)[||] that we can assume a linear dependence: $E(\Phi) = E_0 - \alpha\Phi$. The Young's modulus thus vanishes at $\Phi_0 = E_0/\alpha$. Because of the order-of-magnitude difference between the actin mesh size ($\leq 0.1$ μm) and the size of the minimal shape that we observe ($\geq 1$ μm), we can use a continuum description of the actin cytoskeleton. We also assume that the actin is homogenous and does not redistribute. Electron microscope pictures indicate no tendency of the actin to redistribute. Such a spatial inhomogeneity will not affect the initial instability, although it may contribute to the higher order, nonlinear effects.

**Pearling.** To understand the origin of pearling in the tubular extensions, we consider a single tube under the combined influence of tension and rigidity. Both a thin elastic shell and a solid rod are considered; we will see that the experimental case corresponds to the solid rod. Using a linear approximation, we consider the stability of a tube subject to a peristaltic perturbation of its radius, $r(z)$, of the form $r(z) = R_0 + u \sin(qz) - (u^2/4R_0)$, where $R_0$ is the tube radius, $q$ is the

---

[||]In fact, cooperative effects cause the Young modulus to vanish at the percolation point, which is expected to occur at volume fractions somewhat below $\Phi_0$. Percolation effects are expected to be significant only in a narrow vicinity around the saturation point, where the network is totally disrupted. This cannot be detected within our experimental precision. Saturation of the pearling wavelength may be related to the vanishing of the shear rigidity at a finite concentration of drug.

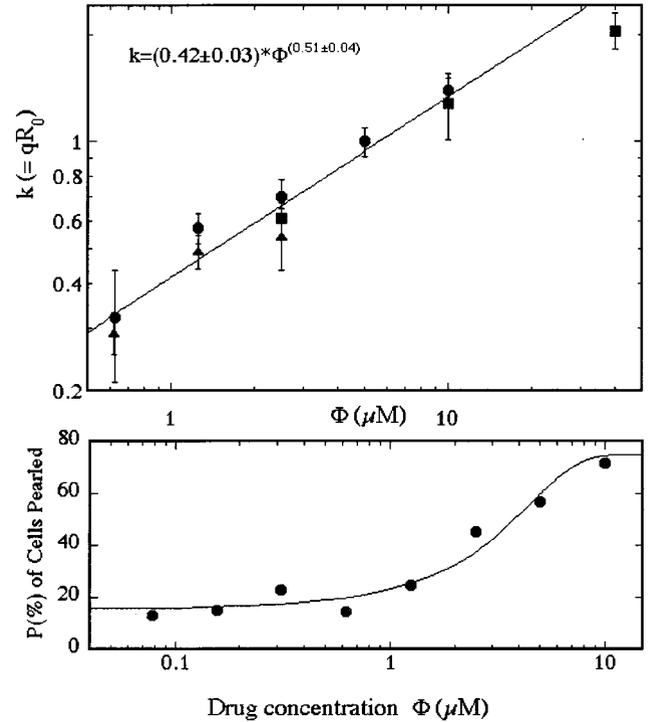

FIG. 3. (*Upper*) Nondimensional wave number $k = 2\pi R_0/\lambda$ of the pearled state as function of the LatA concentration $\Phi$. Samples were scanned by eye to identify cells with pearled tubular protrusions, and still video pictures were transferred to the computer for analysis. The diameter of pearls $2R$, distances between them $\lambda$, and the diameter of tube sections connecting pearls $\delta$ were measured. In many of the cells for which we identified pearling, the instability was in the nonlinear state with a periodic array of isolated pearls rather than a small-amplitude sinusoidal modulation. In such cases, the wavelength was taken as the distance between pearls. To determine the dimensionless wavenumber of the instability $2\pi R_0/\lambda$, we needed to know $R_0$, the initial unperturbed tube radius, which was unknown for fixed cells. Hence, we used volume conservation along the tube: $k = 2\pi R_0/\lambda = 2\pi(4R^3/3\lambda^3 + \delta^2/\lambda^2 - 2R\delta^2/\lambda^3)^{1/2}$. For each drug concentration, we averaged up to 20 tubes and repeated the measurements at least once for almost all drug concentrations, with good reproducibility in each case. At low concentrations ($\Phi \leq 0.5$ μM) the wavelength measurements are noisy and therefore unreliable. At high concentrations, a trend to saturation of the wavenumber begins to be observed. Accordingly, we do not present the very low $\Phi$ points and did not consider them or the highest $\Phi$ point in fitting the power law. (*Lower*) Percentage of cells in which at least one cylinder has undergone pearling as a function of LatA concentration $\Phi$. About 100 cells were counted per drug concentration. The line is a fit to an error function centered around $\Phi_c = 2.5$ μM and with a width $s = 3$ μM. This is an integral of the Gaussian distribution $erf(\Phi) = C(2\pi s^2)^{-1/2} \int_{-\infty}^{\Phi} exp[(\Phi' - \Phi_c)^2/2s^2]d\Phi'$. A fraction of about $(1 - C) = 25\%$ of the cells do not pearl even at high $\Phi$.

wavenumber of the modulation, and $u \ll R_0$ is its amplitude. Volume conservation is ensured by the term $u^2/4R_0$. (This assumes that the cell does not exchange volume through its membrane on the time scales involved, and that osmotic effects are negligible. Note that the depolymerization of actin can only produce a small perturbation on the overall osmotic pressure. The addition of monomers would in any case tend to increase the volume of the cell, contrary to our observations.) The area of the tube $S$ is reduced by a factor of

$$\frac{\delta S}{S} = \frac{1}{4}\left(\frac{u}{R_0}\right)^2 \times (k^2 - 1),$$

and so the surface-energy gain, per unit area, is $U_\sigma = \sigma \times (\delta S)/(S)$.

A fluid cylinder under surface tension $\sigma$ is thus unstable to long modes with $k \leq 1$ (8), and it is up to the elastic forces to oppose the instability. In the case of artificial lipids, the bending energy of the fluid bilayer overcomes the instability when the dimensionless ratio of energies is large enough $\kappa/(\sigma R_0^2) \geq 2/3$ (9), where $\kappa$ is the bending modulus of the lipid (28). In cells, the elastic restoring force is mainly caused by the underlying actin cortex and is much stronger.

For the simple case of a thin shell with thickness $h \ll R_0$, elastic deformations of a thin shell can be decomposed into two decoupled components. We will call the in-plane deformations stretching, while regarding out-of-plane deformations as bending (in the case of a full, rigid tube, they remain strongly coupled) (29). The stretching is usually much larger than the bending, giving as a criterion for instability that the dimensionless energy ratio $2/(1 - \nu^2) \times Eh/\sigma$ should be larger than 1, where $\nu$ is the Poisson ratio. Note that this stability criterion does not depend on the tube radius $R_0$.

As mentioned before, thinner tubes are unstable whereas thicker ones are not, so that stability does depend on $R_0$. We conclude that the thickness of the actin layer is close to the tube radius. For a solid rod ($h = R_0$), stretching and bending couple to give the total elastic energy per unit area

$$U_T = \frac{1}{4} \mathcal{E} \left(\frac{u}{R_0}\right)^2 \left(1 + \frac{1}{6} k^2 - \frac{1}{144} k^4\right),$$

where $\mathcal{E} = 3ER_0/1 + \nu$ is an effective cylindrical stretching modulus (30). The exact cylindrical stress function adapted to our boundary conditions is $\chi(r, z) = (\mathcal{E}u/3k^2 I_1(qR_0)) \times J_0(iqr) \sin(qz)$ (31). Expanding in $k$ yields the series $U_T$. Compared to the full numerical solution, we find that keeping terms to second order gives an accuracy that is better than 0.1% for $k \leq 1$, while taking only the constant term gives an error of 15% at most. For all relevant wavelengths ($k \leq 1$), the $k$-dependent terms may be neglected (as can easily be verified numerically).

The new dimensionless criterion for stability is $- U_T/U_\sigma(k = 0) = \mathcal{E}/\sigma \geq 1$, which does depend on $R_0$. It therefore is easier to destabilize thin tubes, requiring less drug and less disruption of the actin cytoskeleton. Because tubes are typically unstable when $R_0 \leq 1$–$2\ \mu$m, and using $R_0 \approx h$, we get an estimate for the thickness of the actin cortex $h \leq 1\ \mu$m, in agreement with published values (10, 11). This also gives an estimate for the average Young's modulus at which pearling occurs, $E_p = (1 + \nu)/3 \times \sigma/R_0 \approx 20$ Pa. As we shall see below, this rigidity is significantly lower than that of an untreated cell.

It is tempting to conjecture that a similar mechanism determines the minimal thickness of the adherent cell without drugs. The threshold for pearling occurs when the rigidity is below a critical value corresponding to a critical concentration $\Phi_c$. Typically, $\Phi_c$ is on the order of 3 $\mu$M but depends on the statistical distribution of tensions (and possibly actin rigidity) in our sample of cells.

Within a linear theory, we can find the growth rate of the instability by considering a small harmonic perturbation to a full rod $u(z,t) = ue^{\omega t} \times \sin(qz)$. The time rate of change of the surface and elastic energies per unit length is $\dot{U} = \pi\omega(u^2/R_0) \times [\sigma(k^2 - 1) + \mathcal{E}]$. The first term represents the tension energy, whereas the second term is the positive shear rigidity contribution that tends to stabilize the tube.

The growth rate of the instability is determined by equating $\dot{U}$ with the viscous dissipation of the Poiseuille flow of water in the deforming tube (32), $W = \eta(16\pi^3/5) \times \omega^2 k^{-2} \times u^2$, where $\eta$ is the cytoplasm viscosity. The resulting dispersion relation takes the form

$$\omega(k) = \frac{5}{16\pi^2} \times \frac{k^2}{\eta R_0} [\sigma(k^2 - 1) + \mathcal{E}].$$

The typical time scale is determined by the growth of the fastest mode ($k = 1/2$), $\tau = (64\pi^2/5 \times \eta R_0/\sigma)$. We can use the typical experimental time scale for development of the instability $\tau \approx 3 \times 10^2$ sec to obtain an estimate for the cytoplasmatic viscosity, $\eta \approx 10^3$ poise. Experimental measurements of the viscosity of the cytoplasm are still controversial, and our estimate tends to support those of ref. 21 rather than those of ref. 24.

Moreover, we find that the typical wavenumber of pearling at criticality is

$$k_c = \sqrt{\frac{\sigma - \mathcal{E}}{\sigma}} = \sqrt{\frac{\Phi - \Phi_c}{\Phi_0 - \Phi_c}}.$$

Comparing to the experiment gives not only the correct square-root behavior, but the measured prefactor also gives an estimate for $\Phi_0$ on the order of 10 $\mu$M. This simple theory seems to reasonably estimate the value at which the actin has lost almost all efficacy. However, the simple linear-dispersion relation cannot predict the full nonlinear dynamic relation that is measured experimentally. The linear theory does not explain why short wavelengths $k > 1$ are observed, but this is known to be characteristic of the nonlinear regime of pearling. Furthermore, the assumption of a linear dependence of the shear rigidity on the concentration may be unrealistic at some limit. The experimental power law does not show the existence of a (radius-dependent) critical concentration $\Phi_c$. This can be attributed to the distribution of tensions and rigidities, which can be high enough to cause pearling at $\Phi < \Phi_c$ and even in untreated cells. Surprisingly, the wavenumber goes to 0 exactly at the limit of untreated cells, hinting at a mechanism of "marginal" stability in untreated cells, which keeps the rigidity in thin tubular protrusions regulated on the verge of instability.

**Cell Shape.** The same physical considerations used to understand pearling can also explain specific features of normal cell shape and their changes on gradual disruption of the actin cytoskeleton. Our simplified model regards the cell as an envelope that is attached to the substrate at several discrete locations and can mediate tension, along with a thin actin cortex shell supplying rigidity beneath it. The balance of forces in the cell will then depend drastically on the geometry. In its initial natural state, the adhering cell has a flattened shape, which is a stable configuration. Most of the elastic energy is located around the edge, giving an effective line tension $\gamma$. Approximating the cross section of the elastic edge as $\pi h^2$ gives $\gamma \simeq \pi E h^2$.

Fig. 4 depicts the calculated equilibrium shapes of the cell for decreasing $\gamma$. Minimizing analytically the total energy $\gamma \int dl + \sigma \int dS$ (elastic line tension on the edge together with the surface tension on the whole cell) gives the force–balance equation $\sigma = \gamma/R_*$. Volume constraints are supposedly taken care of by adjustment of the cell's vertical profile. The equilibrium shape describes a positive-curvature circular arc between the adhesion points, with radius $R_* = \gamma/\sigma \approx \pi Eh^2/\sigma$. Approximating for the untreated cell $R_*(\Phi = 0) = 10\mu$m and $h \approx 1\ \mu$m, whiles taking for the tension values typically measured in cells $\sigma = 4 \times 10^{-2}$ erg/cm$^2$ (19–21), we obtain an estimate of $E_0 \approx 100$ Pa, consistent with values reported in refs. 24–27 (compared with $Ep \approx 20$ Pa for pearling).

As seen in both experiment and calculation $R_* \approx E_0 - \alpha\Phi$ indeed decreases with $\Phi$. At high values of $\Phi$, $R_*$ decreases to a value smaller than the typical distance between adhesion points, and the tubular shape begins to become apparent (Fig. 1 $d$ and $h$ and the theoretical shapes, Fig. 4 $c$ and $g$). The shape of the membrane near the cell center can be shown analytically to be close to a semicircle, i.e., the distance between rays at the contact to the cell center is $L \approx 2R_*$. This prediction is well borne out experimentally ($L/R_* = 2$ within 10% on average).

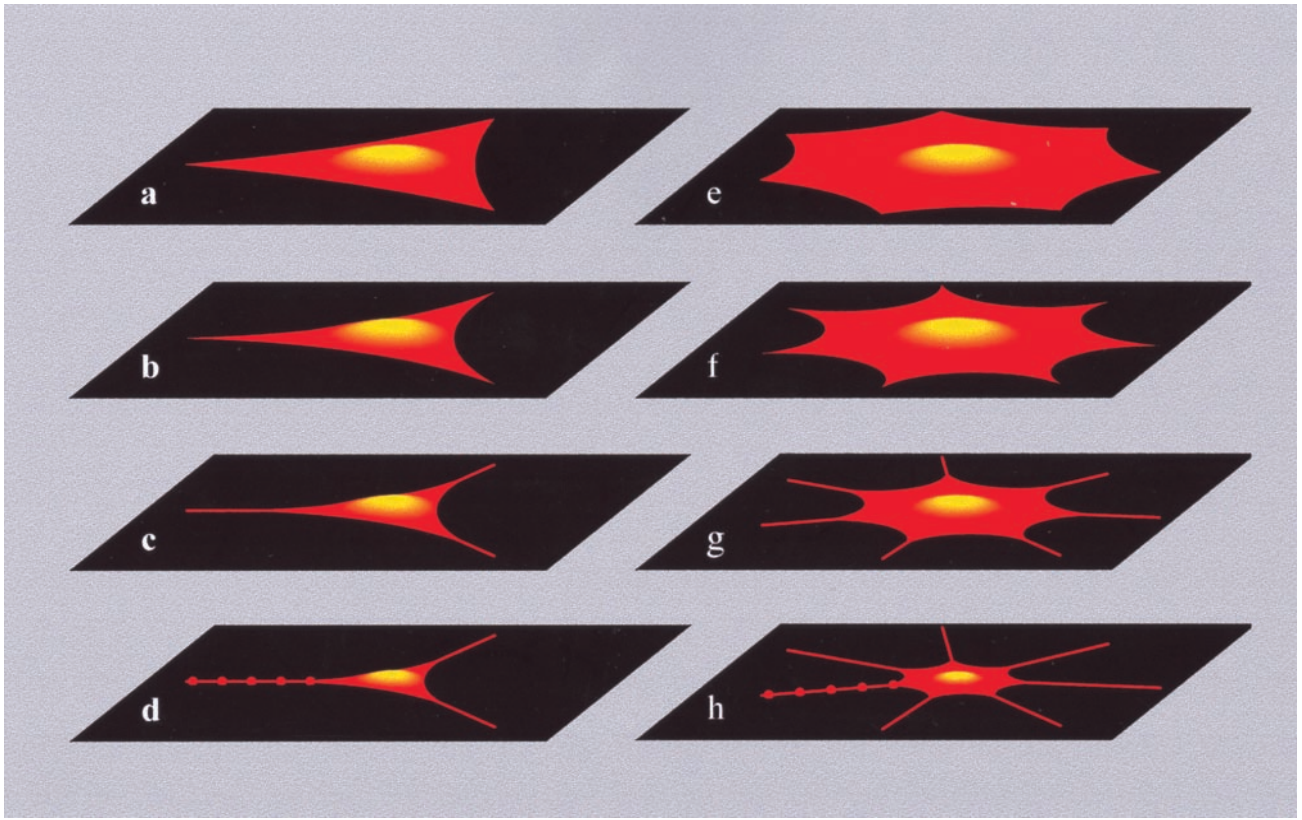

FIG. 4. Shapes of cells calculated from the theory as a function of the changing ratio between the balanced surface tension and effective elastic line tension at their edges. The calculation begins by placing a symmetric polygonal cell with $n$ adhesion points lying on the unit circle ($n = 3$ for $a$–$d$ and $n = 7$ for $e$–$h$). The parameter that determines the shape is then the radius $R_* = \gamma/\sigma$, measured in units of the distance between adhesion points. The values for $a$–$d$ are $R_* = 2.31, 1.15, 0.80,$ and $0.58$ and for $e$–$h$ are $R_* = 1.15, 0.58, 0.40,$ and $0.23$. $a$ and $e$ represent untreated cells, each with different symmetry, determined by the number of adhesion points. In $a$–$d$, we depict a more polar cell with three adhesion points, whereas in $e$–$h$, a more spherical cell like those in Fig. 1 $c$–$e$ is produced. As the concentration of LatA $\Phi$ increases, the radius of curvature $R_*$ that defines the arcs of cell membrane that hang between the adhesion points decreases ($b$ and $f$). Eventually, when $R_*$ is too small for the membrane to bridge the distance between adhesion points, the cell develops cylindrical protrusions ($c$ and $g$). When the actin cortex rigidity of the collapsing cell is decreased below a critical value, these growing protrusions become unstable with respect to the pearling instability ($d$ and $h$). Perspective and the cell height were added for illustration purposes, along with pearls on one cylinder in the final stages ($d$ and $h$).

At higher concentrations, the low surface-to-volume ratio favors a change of sign in the curvature of the outer envelope, the arborization process is complete, and we obtain a "sun" (Fig. 4 $d$ and $h$). This shape has a circular (minimal surface) core containing the nucleus along with most of the cell cytoplasm and rays attached to the original adhesion points.

## Summary

In conclusion, we have shown that the phenomenon of pearling can be quantitatively explained by a simple theory that takes into account the elastic energetics of shape. The energy includes the rigidity of the actin cytoskeleton, which is disrupted by the drug LatA and which competes with a tension produced by external constraints such as adhesion points. The observation of pearling in other systems such as stretched neuronal axons (33–37) can be attributed to the same physical mechanisms. Similarly, because the termini of peripheral axons are very thin, they often demonstrate a pearling morphology (terminal varicosities) (38). The appearance of pearling in untreated cells and the behavior of the wavenumber at low values of concentration $\Phi$ may indicate that the untreated cell regulates its rigidity marginally, just sufficiently to counteract the instability in existing tubular protrusions.

The theory generally can account for cell shape in all of those regions that don't adhere, both in untreated cells and in the various shapes that appear as the cytoskeleton is disrupted. Irrespective of other factors that are involved in determining cell shape, the elastic energy considerations that we have raised must always be taken into account when discussing cell morphogenesis.

S.A.S. acknowledges a grant to the Research Center for Self Assembly from the Israel Science Foundation. E.M. acknowledges support from the US-Israel Binational Foundation Grant 94-00190. A.B. acknowledges the support of La Fondation Rapael et Regina Levy. S.A.S., E.M., and A.B. acknowledge support from the Minerva Foundation, Munich, Germany.